# Silent Data Corruption by 10× Test Escapes Threatens Reliable Computing

Subhasish Mitra[1], Subho Banerjee, Martin Dixon, Mike Fuller, Rama Govindaraju[2], Peter Hochschild, Eric X. Liu, Bharath Parthasarathy, Parthasarathy Ranganathan

Google

# Abstract

*Too many defective compute chips are escaping today's manufacturing tests – at least an order of magnitude more than industrial targets across all compute chip types in data centers. Silent data corruptions (SDCs) caused by test escapes, when left unaddressed, pose a major threat to reliable computing. We present a three-pronged approach outlining future directions for overcoming test escapes: (a) Quick diagnosis of defective chips directly from system-level incorrect behaviors. Such diagnosis is critical for gaining insights into why so many defective chips escape existing manufacturing testing. (b) In-field detection of defective chips. (c) New test experiments to understand the effectiveness of new techniques for detecting defective chips. These experiments must overcome the drawbacks and pitfalls of previous industrial test experiments and case studies.*

# 1. Introduction

Software engineers typically assume that commodity compute chip hardware — e.g., CPU, GPU, Machine Learning (*ML*) accelerator — always works correctly during its lifetime. In digital systems, *hardware errors* – also referred to as *errors* in this paper – occur when incorrect logic values appear on signals in the underlying hardware. Large-scale distributed systems in data centers often assume that hardware errors, if any, immediately cause software execution to stop via crashes and hangs, for example – the *fail stop failure* assumption. Unfortunately, none of these assumptions hold today.

We show that errors caused by test escapes are a major concern today. *Test escapes* are chips with manufacturing defects[3] that today's manufacturing testing fails to detect. Current manufacturing testing practices employ two types of test patterns to detect defective chips: (a) *scan test patterns* (see Sec.

---

[1] Work done at Google. The author's primary affiliation is Stanford University. Email: subh@stanford.edu
[2] Work done when the author was at Google. The author's current affiliation is NVIDIA.
[3] Manufacturing defects are introduced during manufacturing. They are different from design bugs. See [Lin 14] for details about design bugs.

3.2.1) applied via scan Design for Testability (*scan DFT*), and (b) *system-level tests*[4] that use workloads to test chip functionality. These patterns are applied under specific *test conditions* – voltage, frequency, temperature. Our data shows that the number of test escapes encountered in data centers is at least 10-fold higher than published industrial targets. These include test escapes that produce errors right after manufacturing, referred to as *t = 0 defects*, and those that degrade in the field leading to errors, referred to as *t > 0 defects* or *Early-Life Failures (ELF)*. End-of-life circuit aging isn't the expected cause of ELF – we usually encounter ELF much earlier than expected chip lifetimes.

To put this in a broader context, around the early 2010's, various error causes, largely benign in the past, started becoming prominent – radiation-induced soft errors, supply voltage variations, and circuit aging were considered the main challenges. A variety of techniques was created to overcome these challenges. Their adoption by industry is growing rapidly, in markets ranging from data centers to automotive systems. References [Cheng 18, 25] provide an overview of these techniques. Test escapes were not considered as a significant cause of errors at that time.

Hardware errors cause various *system-level incorrect behaviors* [Cheng 18]. Crashes or hangs are common examples. An especially insidious one is called *Silent Data Corruption* or *SDC* – the application terminates normally without any error indication but, at the end of its execution, the application outputs are different from an error-free run. The concept of SDCs is not new. It is represented in the fault-tolerant computing literature as *undetected errors*, *data integrity*, and *fault-secure behavior* (e.g., [Kraft 81]). A defective chip that produces SDCs can crash or hang as well. A machine with a defective chip is *SDC-causing* if we can identify a (bug-free) workload, e.g., a system-level test or an actual application, that produces SDCs when executed on that machine**.** The corresponding defective chip is referred to as an *SDC-causing chip*. While errors produced by SDC-causing chips are of recidivist nature, SDCs are not necessarily produced every time the specific workload is executed on an SDC-causing chip. This is because the internal state and the electrical conditions (voltage, frequency, temperature) can vary each time the workload is executed. Our data shows that test escapes, both t = 0 defects and ELF, result in significant SDC-causing chips. Several hyperscalers (Alibaba, Amazon, Google, Meta, Microsoft) and hardware companies (AMD, ARM, Intel, NVIDIA) have discussed test escapes and SDCs [Dixit 21, Hochschild 21, Parthasarathy 24, Trock 24, Wang 23].

Test escapes and SDCs are a critical threat for several reasons:
(a) Compute chips are *the* backbone of computing infrastructure. When they fail to meet expected reliability standards, the consequences are far-reaching.
(b) Significantly increased cost to develop, maintain, and operate large-scale computing infrastructure:
   (i) The burden on software engineers to debug perceived bugs in their code (when the real culprit is defective hardware) or to make their software resilient to defective hardware is untenable.
   (ii) Detecting, triaging, and removing defective chips from serving systems is expensive.

---

[4] System-level tests are also called functional tests. However, there is a subtle difference between the two. Functional testing of an ML accelerator, for example, might refer to tests that run workloads that fit on a single chip. In contrast, system-level tests might run training or inference spanning multiple chips.

- (iii) Recovering lost and corrupted data is costly, as is handling downtimes and service disruptions.
- (c) Damage amplification:
    - (i) With increasingly distributed systems, a defective chip in one machine can corrupt computations on other machines.
    - (ii) Corruption of critical components, such as security keys or metadata, can result in inaccessible data and services.
    - (iii) With the emergence of Artificial Intelligence/Machine Learning (*AI/ML*), cloud, and parallel computing, a defect in one chip can cause significant waste in computing resources.

Existing fault-tolerant computing practices aren't sufficient to overcome test escapes and SDCs. Safety- and mission-critical systems often use extensive redundancy at the chip and system levels. Examples include duplication, logic parity, and other coding techniques [Mitra 00] inside chips and Dual-Modular-Redundancy (*DMR*), Triple-Modular-Redundancy (*TMR*), and Byzantine Fault Tolerance at the system level. Such approaches incur large energy, execution time, and area overheads for commodity compute hardware. Unlike errors that stem from radiation, circuit aging or supply voltage variations, smart techniques to overcome errors caused by test escapes are in their infancy today.

This paper is a call to overcome the challenge of test escapes. In Sec. 2, we present our data and observations highlighting the severity of test escapes and SDCs caused by them. Sec. 2 also shows that existing manufacturing testing practices aren't making enough progress in reducing test escapes to the required levels. Hence, in Sec. 3, we present a three-pronged approach to future directions in overcoming test escapes and SDCs caused by t=0 defects and ELF. The core components of our three-pronged approach are:

1. **Quick diagnosis of defective chips directly from system-level incorrect behaviors.** Such diagnosis is critical for gaining insights into why so many defective chips escape existing manufacturing testing (details in Sec. 3.1).
2. **In-field detection of defective chips.** By in-field, we mean detection of defective chips that pass manufacturing testing, both prior to and after deployment in data centers (details in Sec. 3.2).
3. **New test experiments to deeply understand the effectiveness of new techniques for detecting defective chips.** These new test experiments must overcome the drawbacks and pitfalls of previous industrial test experiments and case studies (details in Sec. 3.3).

# 2. Data and observations

The turmoil caused by test escapes came to our attention in 2016/2017. Our large computing fleet helped us obtain statistically significant data. Subsequently we invested significant efforts in creating detection techniques, driving root-cause analysis in collaboration with our vendors, and improving the resilience of our critical services. Conclusively proving if a chip is defective or not by running applications and system-level tests in the field is hugely complex, challenging, and time-consuming, requiring deep human expertise for accurate triaging over several months.

## Observation 1: The number of test escapes is very high, around 5,000 defective parts per million (DPM) over lifetime, irrespective of the compute chip type. These include both t=0 defects and ELF.

In contrast, the industry expects test escapes to be at the level of 100-500 DPM or less over lifetime for commodity compute chips. These include all manufacturing defects – including ELF – not just those resulting in SDC-causing chips.

Defective chips that pass manufacturing testing are detected in-field using various techniques such as system-level tests, kernel crashes, application-level checks, uncorrectable machine checks from residue/parity errors in arithmetic units and address/data buses, uncorrectable errors in on-chip memory, and hardware watchdog timeouts. Many such chips result in swaps; i.e., the chips are replaced by the vendor. We estimate that test escapes cause 0.5% – i.e., 5,000 DPM – of chips to be swapped. Other reasons for chip swaps, e.g., hardware design bugs and analog circuit issues, aren't included in this estimate. Most swapped chips aren't fully diagnosed because the economics of managing and triaging is prohibitive — simply replacing them without deeper analysis is unfortunately more practical from a business perspective.

## Observation 2: Test escapes resulting in SDC-causing chips are estimated to be around 1,000 DPM. SDC-causing chips are a challenge irrespective of the compute chip type.

A subset of test escapes resulting in chip swaps is further analyzed using extensive system-level testing for possible SDC-causing behaviors. Chips exhibiting SDC-causing behaviors are then flagged as SDC-causing. Table 1 summarizes SDC-causing chips experienced in the Google fleet over 4 generations of computing platforms, spanning multiple vendors across 22nm to 5nm process technologies. Table 1 should be viewed with care since SDC-causing chips corresponding to a platform are normalized by that platform's total volume – the older-generation platforms have been in production for a longer period of time. Testing capabilities have also improved over time.

The conclusion from Table 1 is: SDCs caused by test escapes continue to be a major challenge across all computing platforms. While recent generations undergo more manufacturing testing, the challenge hasn't materially improved. For some platforms, there may be fewer SDC-causing chips because the corresponding test content is not mature or thorough enough. Given the thoroughness challenges of system-level testing (detailed in Sec. 3.2.1), Table 1 provides a lower bound on the number of SDC-causing chips.

| Platform | Generation | Test escapes resulting in SDC-causing chips (DPM) |
|---|---|---|
| Platform 1 | 1 | 318 |
| Platform 2 | 1 | 1175 |
| Platform 3 | 1 | 653 |
| Platform 4 | 2 | 1097 |
| Platform 5 | 2 | 794 |
| Platform 6 | 2 | 1605 |
| Platform 7 | 3 | 1096 |
| Platform 8 | 3 | 495 |
| Platform 9 | 3 | 345 |
| Platform 10 | 3 | 1857 |
| Platform 11 | 4 | 625 |

**Table 1**: Test escapes resulting in SDC-causing chips across 11 different platforms. (Generation 1 oldest, Generation 4 newest).

## Observation 3: SDC-causing chips produce errors frequently.

We ran a large number of system-level tests to check for incorrect outputs (which when unchecked would cause SDCs) produced by SDC-causing chips in Platforms 2 and 5 in Table 1. We found that the median rate of incorrect outputs produced by these chips is 820K per billion chip hours of checking time (Table 2). These rates are much higher compared to radiation-induced soft errors for example. We have confirmed that incorrect outputs produced by these SDC-causing chips are recidivist in nature. A key consideration is whether these rates, obtained from system-level tests in Table 2, are representative of actual applications. For ML accelerator chips, we inserted fine-grained checks into thousands of training workloads to check 1% of their total runtime and obtained a similar conclusion (Table 3).

|  | Incorrect outputs per billion chip hours of checking time |
|---|---|
| Median | 820K |

**Table 2**: Incorrect outputs from known SDC-causing CPUs.

|  | Incorrect outputs per billion chip hours of checking time |
|---|---|
| Median | 916K |

**Table 3**: Incorrect outputs from known defective ML accelerator chips running training workloads.

## Observation 4: Diagnosis and root-cause analysis of test escapes are very limited.

Detected defective chips are promptly removed from the serving fleet. A small fraction (less than 10%) of "interesting" cases are sent back to vendors for detailed analysis – to localize defects and identify their underlying cause – broadly referred to as diagnosis and root-cause analysis[5]. The purpose is to investigate defects that escape existing manufacturing tests and to improve test content and manufacturing practices. Returned chips aren't chosen randomly: we try to identify chips that we anticipate (in collaboration with the vendor) would be worthy of deeper root-cause analysis. Figure 1 shows the results of such analysis:

(a) **No Trouble Found or NTF (36%)**: These are chips for which the vendor is unable to produce incorrect behaviors – a major challenge. NTFs occur for several reasons:
   (i) Vendors struggle to create incorrect behaviors in their own environments[6], and the behaviors that they can create may or may not match those observed in production systems.
   (ii) Some returned chips may produce incorrect behaviors due to design bugs and not manufacturing defects.
   (iii) When a multi-chip system fails, it may be difficult to identify a single defective chip — as a result, an ensemble of chips (including defect-free chips) might be sent to the vendor for root-cause analysis.
(b) **ELF (29%)**: These chips fail on vendor's test content that previously passed. We estimate that 29% of chips sent for root-cause analysis are caused by ELF.
(c) **Test gap fixed (18%)**: These are defective chips with test gaps. The vendor is able to identify and develop new tests to ensure that chips containing similar defects can be detected.
(d) **Test gap (10%)**: These are defective chips with test gaps that the vendor is aware of but hasn't been able to develop successful tests for similar defects.
(e) **Damaged (7%)**: These chips were damaged when transitioning from the production environment to the vendor's environment. Despite best efforts, this number is unusually high.

---

[5] In testing literature, there is a difference between diagnosis and root-cause analysis (also called failure analysis). To keep things simple, we do not make that distinction in this paper.
[6] For logistical, technical, and legal reasons, vendors are often unable to run full workloads in their testing environments.

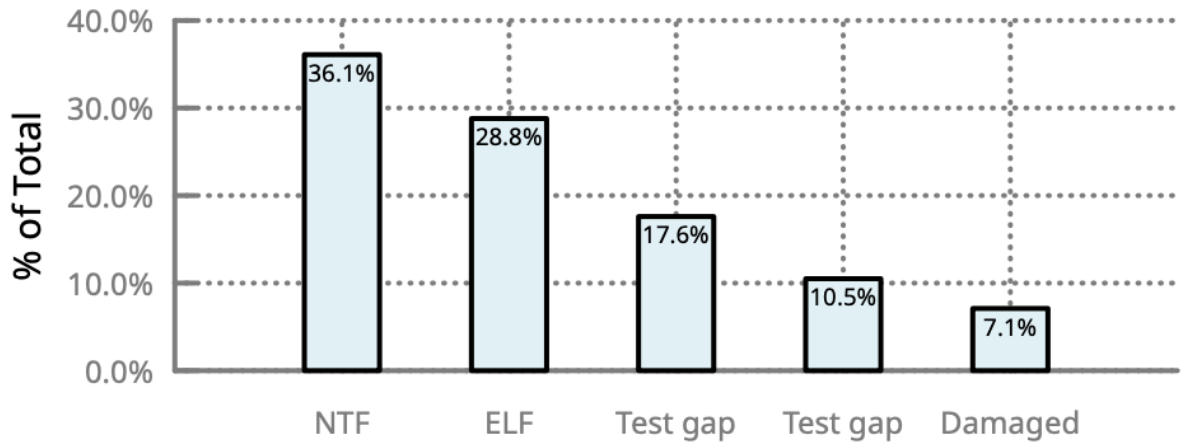

**Figure 1**: Summary of analysis of defective chips returned to vendors for root-cause analysis.

## Observation 5: Most test escapes are detected after deployment in data centers.

Table 4 shows the number of defective machines, i.e., machines containing defective chips, detected using various approaches as part of our server workflow for a server platform deployed in large volume within our fleet.

Pre-deployment testing is performed in two scenarios: (a) after chips are assembled into systems but before the resulting systems are deployed in production, and (b) after the resulting systems are installed in the fleet but before any workloads are scheduled on them. For Table 4, a few days of pre-deployment testing post-installation (i.e., scenario (b)) detected 12% of the defective machines.

Post-deployment, online testing leverages unused cycles during low utilization to run tests without any downtime visible to the end-user. Online testing might run concurrently with other workloads. Offline testing is performed when (a) machines in the fleet are drained and no workloads are scheduled or (b) machines are taken offline for repair due to various known hardware failures such as faulty DIMMs or fans. The downtime cost of draining machines and bringing them offline can be significant. Together, online and offline testing methods detected 29% of defective machines for this example.

Our system health infrastructure uses forensics, identifying potentially defective machines by known signals and suspicious events (details in Sec. 3.2.3). This infrastructure helped us detect 49% of defective machines. Finally, user-level detection, which relies on checks built into critical software services, helped detect 10% of defective machines.

Each row in Table 4 reports defective machines that were missed by detection approaches above it. The cost and effort to detect defective machines increase significantly as we go down the rows. Pre-deployment detection is less costly than online/offline testing or user-level detection. Beyond the risks and recovery costs associated with defective chips and SDCs post-deployment, the software development effort also rises sharply. For instance, relying on user-level techniques requires substantial software engineering investment to insert extensive checks. Ideally, all defective chips should be detected by pre-deployment testing, or, preferably, by the vendor during manufacturing testing.

| Detection approach | % defective machines detected | Defective machines detected (per million) |
|---|---:|---:|
| Pre-deployment testing | 12% | 479 |
| Post-deployment | | |
| Online/Offline testing | 29% | 1099 |
| System health and forensics | 49% | 1886 |
| User-level | 10% | 393 |

**Table 4**: Most defective machines are detected post-deployment.

# 3. Call to action: a three-pronged approach to overcome test escapes and SDCs

As discussed in Sec. 2, we aren't seeing enough progress to rely on existing manufacturing testing practices alone for drastically fewer test escapes – both t=0 defects and ELF. The following factors make existing manufacturing testing difficult:
   (a) Complex manufacturing floor economics impose severe test time constraints. Exceeding several minutes of per-chip test time is often regarded as impractical. The infrastructure required to apply long system-level tests, lasting up to several hours, is considered infeasible during manufacturing.
   (b) Today's manufacturing test practices lack ways to accurately estimate outgoing test escapes. This is evident from the large number of test escapes reported in Sec. 2. There are instances where test content used by vendors and hyperscalers is too similar, which further challenges test escape estimation. In addition, there is a severe lack of robust feedback mechanisms, resulting in a lack of insights into why a large number of defective chips escape existing manufacturing testing (Sec. 2, Observation 4).
   (c) As reported in Sec. 2 (Observation 4), a significant fraction of test escapes are caused by ELF despite the use of hardware burn-in through voltage and temperature stress during manufacturing testing.

Hence, we must create alternate methods to detect defective chips, especially with the high prevalence of ELF. We have outlined a three-pronged approach for this purpose:
   (a) Quick diagnosis of defective chips directly from system-level incorrect behaviors, enabling deep understanding of causes of test escapes (Sec. 3.1).
   (b) In-field detection of defective chips, both t=0 defects and ELF (Sec. 3.2).

(c) New test experiments that overcome the drawbacks and pitfalls of previous industrial experiments and case studies (Sec. 3.3).

In the rest of this paper, we discuss technical challenges and research directions for this three-pronged approach.

## 3.1. Quick diagnosis directly from system-level incorrect behaviors

There is an urgent need for new ways of diagnosing defective chips quickly and directly from system-level tests and workloads in the field – without relying on special testing environments – for obtaining meaningful insights into why large numbers of defective chips escape existing manufacturing testing. Existing diagnosis of field returns requires several months of intense effort by many engineers and often leads to unsuccessful outcomes (Sec. 2, Observation 4).

The in-field aspect – i.e., no reliance on special testing environments – is critical. As discussed in Sec. 2, vendors are often unable to run full workloads in their testing environments. Instead, they often re-run their manufacturing tests (or extended versions of those tests) over extended test conditions (typically not used for manufacturing testing). Defective chips often pass these tests resulting in NTFs. Aggressive test conditions can make defect-free chips fail, which don't provide insights into defective chips.

For quick diagnosis directly from system-level incorrect behaviors, the following questions must be addressed:

(a) **How do we quickly differentiate hardware defects vs. software bugs from system-level incorrect behaviors?** This is a daunting task and today's approaches are very time consuming, requiring multiple engineering teams spanning different systems that make up the distributed infrastructure. We envision new approaches through a combination of special hardware checks inserted at various parts of a design and new analysis techniques that exploit the inserted hardware. As an extreme design point, inserting duplication-based error detection in hardware significantly simplifies this challenge but has high area and energy costs.

(b) **How do we localize hardware defects directly from system-level incorrect behaviors?** Defects must be localized to small enough regions in hardware, together with short input sequences that activate those defects, for the triaging team to understand and determine root cause. The following are the key challenges in performing such localization:
  (i) Error detection latency, the time elapsed between the occurrence of an error caused by a hardware defect and its manifestation as an observable incorrect behavior at the system/application level (e.g., crash, hang, or detection by software-instrumented checkers), can be extremely long – billions of clock cycles [Lin 14]. It is impractical to trace back too far in system execution history for diagnosis and root cause analysis. Techniques to drastically shorten error detection latencies, e.g., [Lin 14], are required.
  (ii) Reproducing system-level incorrect behaviors is very difficult due to asynchronous I/Os and interrupts, execution order entropy arising from locks/synchronization, internal states of the system not visible to software, and voltage/temperature effects. Depending

(iii) System-level simulation is often used to obtain the golden response of a (defect-free) chip, including internal states not visible at the software level. Such simulations are orders of magnitude slower than actual silicon. Therefore, (cycle-accurate) simulation to obtain expected values of internal states is next to impossible except for the shortest system-level tests – in contrast, today's system-level tests that detect defects can be extremely long.

on the system-level test infrastructure, the input data may not be identical across multiple runs of the same test. Moreover, it might take a very long time (even days) for a test to detect errors.

Continuing with the duplication-based extreme design point we discussed in part (a) above, hardware duplication and checkers at the granularity of combinational logic blocks can overcome many of these challenges but at exorbitantly high area, energy, and power costs [Cheng 18, Mitra 00]. Hence, new low-cost approaches are essential.

## 3.2. In-field detection of defective chips: t=0 defects and ELF

We focus on two complementary approaches for in-field detection of defective chips:
(a) In-field testing (Sec 3.2.1): Special test patterns and test conditions are applied in-field during pre-deployment and post-deployment testing. The economics of test times for in-field testing allows far longer test times compared to typical manufacturing testing. These longer test times create opportunities for more extensive test patterns and test conditions far beyond existing manufacturing testing. Furthermore, in-field testing also addresses ELF.
(b) Error detection in the field (Sec. 3.2.2): Defective chips are identified by detecting errors produced by them as they execute workloads in the field.

In-field testing and error detection can be proactive or reactive [Li 09]; i.e., these mechanisms may be invoked periodically or based on events such as those triggered by system health and forensics (Sec 3.2.3).

## 3.2.1. In-field testing

### In-field system-level testing

System-level tests, lasting up to several hours, are the workhorse for in-field testing today. Existing approaches for creating system-level tests, including those based on random instructions and fuzzing, are *ad hoc* and lack solid technical foundations. While system-level tests detect substantial test escapes, as presented in Sec. 2, their thoroughness is unclear. We encountered many situations where these tests seemed comprehensive – evidenced by a declining rate of test escape detections over time – only to discover numerous additional escapes when we performed further system-level testing. Moreover, the impact of test conditions on system-level testing is not well understood. There are potential research opportunities for systematic ways to create system-level tests for domain-specific hardware accelerators. Examples include thorough functional tests for systolic arrays [Cheng 87] and accurate high-level modeling of hardware errors for specific accelerator architectures [He 23].

## In-field scan testing

Unlike system-level tests, scan DFT offers the potential for rigorous and principled approaches to in-field testing by automatically generating thorough test patterns that have quantified coverage with respect to meaningful metrics. The effectiveness of scan DFT in detecting today's test escapes is sometimes debated. However, there is little evidence that scan testing, when combined with proper test patterns and conditions, is incapable of detecting defective chips that escape existing manufacturing tests.

One promising approach for scan DFT-based in-field testing is CASP – Concurrent Autonomous chip self-test using Stored test Patterns [Li 08, 13]. CASP and its derivatives have recently been implemented in several industrial products. Examples include Amazon's Deterministic In-Fleet Scan Test [Trock 24], Intel's In-Field Scan [Inkley 24], NVIDIA's In-System Test [Jagannadha 19] and several others. The main CASP ideas are given below:
  (a) CASP stores highly thorough and compressed test patterns in off-chip storage such as FLASH. The stored test patterns can be updated according to application reliability constraints and failure characteristics of chips discovered over time.
  (b) CASP provides special hardware/software support to (i) fetch scan test patterns and expected responses from off-chip storage quickly using high-speed I/Os (unlike slow test data transfer between the chip and the tester during manufacturing testing, limited by the frequency of scan chain shifts), (ii) apply the fetched test patterns to one or more components inside the chip using scan chains inside those components, and (iii) ensure no downtime is visible to the end-user when CASP is operated in online testing mode. CASP can also be operated in offline testing mode.
  (c) CASP provides flexible trade-offs across a wide range of design points. Examples include: (i) balancing trade-offs between scan test data transfer time, scan test application time and area overheads of scan DFT and test compression for CASP, (ii) simultaneous scheduling of application tasks and scan tests when CASP is operated in online testing mode, and (iii) the interplay between hardware, operating system and virtualization support for CASP orchestration. [Li 13] shows CASP overheads to be small: 1% area cost, 1% power cost, and 3% performance cost (when CASP is invoked in online testing mode). Potential security vulnerabilities of CASP in-field testing are also important.

CASP exploits advances in test compression, the rise of parallel and many-core architectures with the end of Dennard scaling, the emergence of inexpensive and high-density off-chip FLASH storage, and the pervasive use of high-speed and high-bandwidth I/Os for transferring data between off-chip memory and parallel computing engines on-chip. CASP avoids key limitations of prior in-field testing approaches (details in [Li 13]) such as pseudo-random logic Built-In Self-Test (BIST) which suffers from long test times, low coverage across various test metrics, inability to handle unknown logic values (X's), and inflexibility in updating test patterns post-manufacture.

While CASP and its derivatives provide the hardware/software foundation for thorough scan testing in the field, key questions remain: what scan test patterns and test conditions should be applied in-field?

Scan test patterns are generated using Automatic Test Pattern Generation (*ATPG*) tools targeting various test metrics such as stuck-at, transition, small delay defect, and cell-aware fault coverage. As Sec. 2 indicates, these metrics, together with test conditions applied during today's scan testing, are inadequate in detecting sufficient levels of defective chips. There are several causes of this situation:
   (a) As discussed in Sec. 3.1, little information is available about why defective chips escape existing manufacturing testing. However, various hypotheses, such as marginal defects as potential causes of test escapes, have emerged. The effectiveness of scan testing in detecting test escapes has also been questioned. [Ryan 14] defines marginal defects as follows: "a marginal unit fails (or will in the future fail, or intermittently fails) at some voltage-temperature-frequency set point within the product's specified operating envelope, while successfully operating at other set points within the specified envelope." [Ryan 14] provides a manufacturing viewpoint by acknowledging that "testing the whole voltage-frequency-temperature operating envelope with comprehensive test content quickly becomes an intractable problem from test time and cost perspectives."
   (b) Faulty circuit behaviors assumed by many test metrics don't match actual defects. Stuck-at faults are known to be grossly inaccurate [McCluskey 00]. Metrics such as cell-aware fault coverage [Hapke 11] rely on defect distributions, e.g., impedances of opens and shorts caused by defects. Such distributions may be difficult to obtain or may be incomplete. Even if they were available, characterizing their effects at numerous locations inside every circuit under test can quickly become computationally infeasible. Hence, ATPG tools often focus on defects inside library cells only (although defects can occur anywhere inside a chip) and precompute their effects. Given these discrepancies and inaccuracies, today's scan testing mostly detects defective chips fortuitously – for example, for timing-independent combinational (TIC) defects, [Nigh 24] estimates that the detection of over 90% of defective chips cannot be explained by the 1's and 0's that are imposed by the corresponding test metrics.
   (c) Today's ATPG tools rarely analyze test conditions comprehensively, especially for tests targeting timing-dependent defects. For example, the longest path through a logic gate may change depending on the operating voltage. Thus, test patterns for timing-dependent defects that propagate transitions along long paths might have to change depending on operating voltages applied during testing.

For thorough scan-based in-field testing using CASP and its derivatives, the following future directions are essential:
   (a) There is an urgent need for new test metrics that overcome the inaccuracy and fortuitous detection challenges of existing ones. It is critical that:
      (i) These new metrics do not require detailed defect distributions, which may be very difficult to obtain.
      (ii) They are computationally scalable especially by utilizing vast compute resources (e.g., not just CPUs but GPUs as well) in the cloud.

The Pseudo-Exhaustive Physically-Aware Region Testing approach (PEPR [Li 22]) is promising because it satisfies the above criteria. While PEPR might create far more test patterns (e.g., 10×-100× [Nigh 25]), the resulting longer test times can be compatible with the economics of in-field testing. However, for comprehensive detection of test escapes, the PEPR metric, which currently addresses TIC defects, must be extended to target sequence- and timing-dependent defects.
   (b) Test conditions must be smartly chosen. Very little research literature covering this aspect exists. The following three points highlight the deep interplay between various factors:

(i) With no external testers, on-chip circuits are required for voltage and clock control during in-field testing. Existing on-chip power management circuits, on-chip self-tuning to overcome circuit aging, and on-chip DFT and Design-for-Debug circuits may be extended and reused for this purpose (e.g., [Keller 07, Kim 13, Li 09 and others]). This requires comprehensive understanding of the granularity and range of voltage and clock control required for detecting defective chips, design of such circuits and the corresponding area and energy overheads, and the impact of such circuits on ATPG. Control of temperature during in-field testing is tricky, to say the least.
(ii) Test conditions supported by on-chip circuits must be taken into account during ATPG with direct impact on test generation effort, in-field scan test application time, and test pattern storage. References [Kim 13, Wu 25] address some of these points.
(iii) When in-field scan testing is applied in online mode, the impact of test conditions and scan test patterns on other (functioning) parts of a chip must be thoroughly analyzed. Conversely, the effects of activities in other (functioning) parts of a chip on test conditions experienced by the circuits under test must also be thoroughly analyzed and understood.
(c) Beyond ATPG, scan testing has had a tremendous impact on manufacturing yield by enabling effective defect diagnosis. However, existing diagnosis techniques rarely analyze test conditions. New diagnosis techniques that account for test conditions in conjunction with test patterns are required especially for timing-dependent defects.

## 3.2.2. In-field error detection

Error detection forms the basis for fault-tolerant computing. A wide variety of error detection techniques [Cheng 18], from the circuit and logic levels all the way to the application level, can be potentially used to detect errors caused by test escapes. Many of these techniques impose large energy, performance, and area overheads[7]. Hence, the key question is: can we achieve required thoroughness, i.e., detect enough test escapes, at small performance, energy, and area costs?

Error detection costs can be reduced by using one or more of the following strategies [Cheng 18]:
(a) By using low-cost techniques that target the following:
(i) Specific error causes, e.g., special flip-flops targeting radiation-induced soft errors, and special circuit failure prediction that targets circuit aging.
(ii) Specific applications such as matrix operations, signal processing, cryptography, and compression. Such techniques can be attractive for domain-specific hardware accelerators. Application-level assertions, implemented in hardware or software, also belong to this category. However, quantifying their effectiveness requires extreme care because, unfortunately, published results are often inaccurate [Cheng 25].

---

[7] Over 100% area and energy overheads are incurred by logic duplication for the blocks being duplicated and checked. Logic parity prediction overheads are tricky to quantify because the values depend on the amount of sharing between logic circuits driving various outputs of the block being checked. For classical logic parity prediction, the overheads can be high – even close to duplication [Mitra00]. If logic sharing constraints are relaxed, guarantees about detection of errors inside logic circuits no longer hold. Despite that, the area and energy overheads can continue to be high (over 20% in [Cheng 18]).

(b) By selectively protecting only certain hardware blocks based on their criticality. Selective protection requires accurate reliability analysis. For example, for radiation-induced soft errors, the architecture-level criticality of flip-flops is analyzed using detailed flip-flop error injections for a given family of workloads. These error injections use accurate error models that are validated by using hardware experiments.
(c) By invoking error resilience mechanisms only during certain points in time to reduce performance and energy costs, also referred to as sampled checking. For example, circuit failure prediction is invoked sporadically since circuit aging is a slow process.

Error detection techniques targeting test escapes have not been extensively studied in research literature and many open questions exist. We highlight two such questions below.

The first question is: *Can we identify critical hardware blocks with respect to test escapes similar to prior work on identifying critical flip-flops for radiation-induced soft errors?*

There are several challenges associated with this question:
(a) Beyond the concept of critical area [Shen 85], little research exists on quantifying the susceptibility of hardware blocks to manufacturing defects. Extensive characterization of defects that escape existing manufacturing tests is even more difficult (Sec. 3.1).
(b) Propagating errors produced by defects inside hardware blocks to the architecture level to assess architecture-level criticality is highly challenging. It requires fault models that accurately capture the effects of defects at the logic level. Such accurate fault models do not exist for manufacturing defects today (Sec. 3.2.1) unlike radiation-induced soft errors inside flip-flops for which hardware-validated fault models exist.

The second question is: *Are software-only techniques, e.g., those based on instruction re-execution and checking [Lin 14, Oh 02, and others], effective in detecting errors caused by test escapes?*

Naive analysis might conclude that re-executing the same instruction sequence on defective hardware produces the same erroneous results, which might then go undetected. However, as shown in [Lin 14], for complex digital systems, this isn't a major concern due to the presence of internal states. Such techniques can be further boosted by diversity enhancement techniques (e.g., [Lin 14]) to avoid identical errors produced by multiple runs of the same instruction sequence. Implementing re-execution-based error detection inside industrial software requires extreme care – to handle tremendous complexity introduced by concurrency-induced non-determinism and myriad independently-written libraries, to maintain sufficient error detection coverage while ensuring no false positives, and to minimize overall performance impact.

The recidivist nature of errors produced by manufacturing defects (Sec. 2, Observation 3) creates an opportunity to reduce in-field error detection costs: error detection techniques do not need to run continuously (similar to the sampled checking strategy described in (c) above). This is because manufacturing defects, both t=0 defects and ELF, are of permanent nature even though their effects might be more pronounced only at certain voltage, frequency, and temperature conditions. Thus, error detection mechanisms can be invoked via sampled checking, albeit at higher system complexity, to reduce performance and energy overheads.

A spectrum of software-only techniques emerges with diverse trade-offs: from "eventual detection" of defective chips at low cost via sampled checking (i.e., defective chips might produce SDCs before they are eventually detected), all the way to expensive and continuous checking to prevent SDCs produced by defective chips. This approach creates opportunities for new research results on various sampled checking strategies, costs incurred by them, ways to manage system-level complexity introduced by them, probabilistic bounds on defective chip detection coverage, and extensive in-field data collection to quantify their benefits.

### 3.2.3. System health and forensics

A defective chip might produce signals such as kernel crashes, hangs, hardware exceptions, anomalous application process crashes, and suspicious events (e.g., when application-level invariants added by engineers to detect software bugs fire). We refer to these signals as system health and forensics (Table 4). The primary challenge is that such signals can come from multiple sources, including manufacturing defects, hardware design bugs, and software bugs.

In our experience, signals from system software are particularly valuable because these components run everywhere. Moreover, system software is mature, well-hardened, and might have a lower incidence of software bugs. A heuristic that is effective in identifying defective CPUs is Core-Concentrated Kernel Crashes (CCKC). These are kernel crashes that are overly concentrated on a single physical core of a single CPU in a particular machine. For instance, a defective core might be indicted if 80% or more of kernel crashes occur on a single physical core, with at least 5 crashes and 3 or more distinct top-of-stack symbols within a 30-day period. The intuition is that, since defects are generally localized, it is unlikely that several cores are affected by a single defect. Similarly, defective hardware should not produce many crashes on the same kernel function (which is much more likely to be a software bug). Hence, we filter out cases with repeated failures on the same top-of-stack symbols. Once identified using this heuristic rule, these CPUs undergo extensive targeted testing using Google-built and vendor-supplied system-level test diagnostics. We observed that in over 70% of the cases, those cores (originally indicted by CCKC) were SDC-causing. Less than 10% of the CCKC detections resulted in false-positive indictments (i.e., we couldn't observe SDC-causing behavior). The rest had a "Maybe" disposition, where the presence of a defect cannot be determined conclusively without explicit human intervention. Core-concentration of fail-stop failures has also been discussed in [Dutta 25].

### 3.3. New test experiments

In the field of testing, many industrial test experiments and case studies have been published. There is a discrepancy between the conclusions from these experiments and the dire situation of test escapes that the industry faces. These gaps are caused by various factors:
- (a) The *actual* population of test escapes may be severely underestimated because functional and system-level tests, used to identify overall defective chip populations (i.e., ground truth), are often not thorough.

(b) The coverage of scan tests with respect to various test metrics is often "low," limited by DFT features or the effort required to achieve high coverage in industrial settings. Fortuitous detection of defective chips, discussed in Sec. 3.2.1, further complicates matters.
(c) A limited set of test conditions applied during testing, together with limited fault diagnosis and root-cause analysis, often limits a clear understanding of the true benefits of new test patterns targeting new test metrics.

There is an immediate need for well-planned test experiments that embrace these and other learnings from previous test experiments and avoid their pitfalls. Inspiration can be drawn from academic test experiments such as the Murphy and ELF experiments [McCluskey 00]. Pros and cons of these academic test experiments are:

(a) Special chips, designed for the sole purpose of being tested, ensure *extremely thorough* testing across a range of technologies. This level of extremely thorough testing may not be achievable for arbitrary designs.
(b) Exhaustive and super-exhaustive test patterns, together with extensive test conditions, enable accurate identification of defective chips – the gold standard. The use of special designs, as discussed above, ensures the practicality of applying such through test patterns (which can quickly become infeasible for arbitrary designs).
(c) Extensive analysis that uses an ultra-wide range of test approaches and a wide range of test metrics, that can be cost-prohibitive on large and complex industrial designs.
(d) Some of the limitations of the academic test experiments are: (i) relatively small number of *defective* chips constrained by production runs, (ii) relatively smaller and less complex chip sizes compared to industrial products, and (iii) less than ideal access to detailed layouts, constrained by the proprietary nature of standard cell libraries used by fabs.

# 4. Conclusion

Far too many defective compute chips are escaping today's manufacturing tests. Test escapes encountered in data centers exceed industrial targets by at least an order of magnitude – across all compute chip types. These include both t=0 defects and ELF. SDCs caused by test escapes, when left unaddressed, pose a major threat to reliable computing.

Manufacturing test practices aren't advancing fast enough to meet this urgent challenge. Progress is stalled in part because diagnosis of field returns is severely limited, yielding little actionable insight and promoting approaches such as system-level testing that lack solid technical foundations today. Traditional fault-tolerant computing, often proposed as a fallback, isn't a viable solution either – it is too expensive for broad deployment.

This isn't all gloom and doom. Now is the moment to seize a golden opportunity to define the future. The three-pronged approach we outlined is key.

First, diagnosis directly from system-level incorrect behaviors in the field is critical for future progress – robust feedback mechanisms that expose weak spots in today's (and tomorrow's) manufacturing test practices. Such mechanisms are largely non-existent today, leaving the industry effectively blind to the true causes of test escapes.

Second, in-field detection of defective chips using CASP and its derivatives provides the foundation for thorough and principled scan testing beyond the constraints of the manufacturing test floor – enabling longer test times optimized in a system-driven context. However, simply reusing today's scan patterns in the field won't suffice; smarter approaches for generating new scan patterns and defining appropriate scan test conditions are essential. This form of scan testing complements in-field error detection, which can potentially be implemented at low cost by exploiting the recidivist nature of errors produced by test escapes. Combined, these methods collectively address both t=0 defects and ELF. Which approach will ultimately prevail remains to be seen, but both offer valuable paths forward – and may even help strengthen manufacturing test strategies.

Third, new test experiments are essential to validate these ideas. These experiments must be designed with rigor – especially in how new detection methods will interact with system workloads, operating environments, and practical deployment constraints. Without this, promising techniques risk being dismissed or misapplied.

AI/ML is expected to play a prominent role across all these three fronts. One promising application is identifying potentially defective machines by analyzing system-level signals – uncovering "needle-in-a-haystack" patterns beyond existing rule-based methods with significantly less manual effort. Another direction is using data-driven approaches for in-field testing and error detection – deciding when to schedule them, which hardware blocks to target, or what test content to apply – based on various hardware and software events and on-chip sensor data. Accurate diagnosis from system-level incorrect behaviors and insights obtained from well-designed test experiments can potentially generate high-quality training data for such AI/ML. Finally, understanding how test escape-induced errors impact AI/ML workloads at the application level is critical for ensuring end-to-end robustness.

# 5. Acknowledgment

We thank Prof. Shawn Blanton of CMU and Prof. Phil Levis of Stanford for valuable feedback, and Stephanie Morton of Google for editorial support. We acknowledge the use of Large Language Models for stylistic editing.# 6. References

[Cheng 87] W.-T. Cheng and J.H. Patel, "Testing in two-dimensional iterative logic arrays," Computers & Mathematics with Applications, Volume 13, Issues 5–6, pp. 443-454, 1987.

[Cheng 18] E. Cheng, S. Mirkhani, L. Szafaryn, C.-Y. Cher, H. Cho, K. Skadron, M. Stan, K. Lilja, J. A. Abraham, P. Bose and S. Mitra, "Tolerating Soft Errors in Processor Cores Using CLEAR (Cross-Layer17